\documentclass[useAMS,usenatbib,usegraphicx]{mn2e}

\newcommand{\xmm}{\it XMM-Newton}

\def\1705{4U~1705--44}

\usepackage{subfigure}
\bibpunct{(}{)}{;}{a}{}{,}

\title[A relativistically smeared spectrum in \1705]
{A relativistically smeared spectrum in the neutron star X-ray Binary \1705:
Looking at the inner accretion disc with X-ray spectroscopy}

\author[T. Di Salvo et al.]{T. Di Salvo$^{1}$\thanks{E-mail:disalvo@fisica.unipa.it},
A. D'A\'\i$^{1}$, R. Iaria$^{1}$, L. Burderi$^2$, M. Dov\v{c}iak$^{3}$, V. Karas$^{3}$, 
\vspace*{0.3cm}
\\ {\LARGE \textup{G. Matt$^{4}$, A. Papitto$^{2,5}$, S. Piraino$^{6,7}$, A. Riggio$^{2,5}$, 
N.~R. Robba$^{1}$, A. Santangelo$^{7}$} 
\vspace*{0.1cm}
}\\
$^1$Dipartimento di Scienze Fisiche ed Astronomiche,
Universit\`a degli Studi di Palermo, via Archirafi 36 - 90123 Palermo, Italy\\
$^2$Universit\`a degli Studi di Cagliari, Dipartimento
di Fisica, SP Monserrato-Sestu, KM 0.7, 09042 Monserrato, Italy \\
$^{3}$Astronomical Institute, Academy of Sciences of the Czech Republic,
Bocni II 1401a, CZ-14131 Prague, Czech Republic \\
$^4$Dipartimento di Fisica, Universit\`a degli Studi `Roma Tre',
Via della Vasca Navale 84, I-00146 Roma, Italy \\
$^{5}$INAF -- Osservatorio Astronomico di Cagliari, Poggio dei Pini, Strada 54, 
09012 Capoterra (CA), Italy \\
$^{6}$INAF -- IASF di Palermo, via Ugo La Malfa 153, 90146 Palermo, Italy \\
$^{7}$IAAT, University of Tubingen, Sand 1, 72076 Tubingen, Germany  } 

\begin{document}

\date{Accepted 2009 June 10.  Received 2009 May 10; in original form 2009 March 10}

\maketitle

\begin{abstract}
Iron emission lines at $6.4 - 6.97$ keV, identified with fluorescent
K$\alpha$ transitions, are among the strongest discrete features in the 
X-ray band. 
These are therefore one of the most powerful probes to infer the
properties of the plasma in the innermost part of the accretion disc
around a compact object.  In this paper we present 
a recent {\xmm} observation of 
the X-ray burster {\1705}, where we clearly detect a relativistically smeared 
iron line at about 6.7~keV, testifying with high statistical significance 
that the line profile is distorted by high velocity motion in the accretion 
disc. As expected from disc reflection models, we also find a significant 
absorption edge at about 8.3 keV; this feature appears to be smeared, and 
is compatible with being produced in the same region where the iron line is
produced. 
From the line profile we derive the physical parameters of the inner accretion 
disc with large precision. 
The line is identified with the K$\alpha$ transition of highly ionised iron, 
Fe~XXV, the inner disc radius is $R_{in} = 14 \pm 2\; R_g$ (where $R_g$ is the 
Gravitational radius, $GM/c^2$), the emissivity dependence from the disc 
radius is $r^{-2.27 \pm 0.08}$, the inclination angle with respect to the 
line of sight is $i = 39^\circ \pm 1^\circ$.
Finally, the {\xmm} spectrum shows evidences of other low-energy emission lines,
which again appear broad and their profiles are compatible with being produced
in the same region where the iron line is produced.
\end{abstract}


\section{Introduction}

Emission lines at $6.4-6.97$ keV are the strongest discrete features in
the X-ray band, and indeed appear to be ubiquitous in all kind of
accreting compact objects.  These have been observed in Active
Galactic Nuclei (AGN, see e.g.\ Tanaka et al.\ 1995; Fabian et al.\ 
2002, and references therein), in X-ray binaries containing stellar 
mass black holes (see e.g.\ Miller et al.\ 2002; Miller et al.\ 2004) as 
well as in X-ray binaries containing a low magnetised neutron star 
(LMXBs, see e.g.\ Bhattacharyya \& Strohmayer 2007; Cackett et al.\ 2008;
Pandel et al.\ 2008; D'A\'\i\ et al.\ 2009).
These lines are therefore a powerful tool to exploit the physical 
properties of the emitting region that in most cases is quite close to the
compact object. The line profile often appears very smeared and
asymmetric, with red wings extending down to 3 keV or even less. 
In most (if not all) cases this profile has been ascribed to
the Doppler and relativistic effects caused by the high Keplerian 
velocities in the accretion disc which extends deeply in the gravitational 
potential well of the compact object.

In fact (see Matt 2006 for a review, and references therein), 
if an emission line is produced in an accretion disc, which
is illuminated by the main Comptonised continuum coming from a corona,
the high velocities in the disc produce a double peaked 
line profile: the blue-shifted peak arising from matter approaching the 
observer and the red-shifted peak from matter receding. The separation 
in frequency between these two peaks depends 
on the line-of-sight velocity difference, which in turn depends on the
inclination of the disc and on the radius where the emission occurs.
At small radii the Keplerian velocities become relativistic, and this
causes that the intensity of the blue peak apparently exceeds that of
the red peak because of Doppler boosting.
Finally, for very small radii, the Gravitational redshift becomes
important and the entire profile may be shifted to lower energies (see
Fabian et al.\ 1989 and Laor 1991 for detailed modelling of these
profiles).  

Although compelling evidence can been found in literature that iron line
profiles often appear broad and asymmetric, and important diagnostics
of the emitting region have been inferred (see e.g.\ Reynolds, Brenneman, 
\& Garofalo 2005 as a review),
alternative explanations for the line broadening have been suggested.
In fact, it is also possible to obtain broad and asymmetric profiles by 
Comptonisation of
line photons in a medium of moderate optical depth and temperature
of a few keV (e.g.\ Kallman \& White 1989; Brandt \& Matt 1994; 
Misra \& Kembhavi 1998; Rozanska \& Madej 2008; see, however, Reynolds 
\& Wilms 2000 for a discussion of the case of MCG--6--30--15) or with 
other kind of non-thermal Comptonisation (Laurent \& Titarchuk 2007); 
complex absorption may also mimic a very broad relativistic
profile (Miller et al. 2008). Even if the relativistic disc line
is often the most convincing solution, it is clear that more direct
and unambiguous evidences would be highly desirable.



In this paper we present a high statistics and well resolved disc line
profile in a LMXB containing a neutron star, together with other identified 
broad emission features and an iron absorption edge compatible with coming 
from the same disc region. We are therefore able to find the disc parameters 
with unprecedented precision. 

\section{Observation and Results}
{\1705} is a LMXB containing a neutron star. It shows type I bursts
and the typical aperiodic variability usually observed for LMXBs of
the atoll class (Olive, Barret, \& Gierlinski 2003).  Similarly to 
X-ray binaries containing black holes, this source shows 
regularly state transitions: from a high/soft state, where the X-ray 
spectrum is dominated by soft spectral components with typical temperatures 
less than a few keV, to low/hard states where the X-ray spectrum is
dominated by a hard thermal Comptonisation (e.g.\ Barret \& Olive 2002;
Piraino et al. 2007).  A Chandra High Energy Transmission Grating (HETG)
observation of this source performed in 2001 (when the source was in a
soft state) revealed the presence of a broad iron line at $\sim 6.5$
keV which was well fitted by a {\it diskline} profile (Di Salvo et
al. 2005).  However, the {\it diskline} profile 
gave an equally good fit as 
using a simple Gaussian profile, the double peaked profile was not
resolved in the Chandra/HETG spectrum, and hence we could not exclude
other possible origins for the broad iron line profile.  
The relativistic nature of the iron line profile in {\1705} has been very
recently confirmed by a Suzaku observation;
the inferred inner radius of the disc was $R_{in} \sim 10\; R_g$, while 
the inclination of the system with respect to the line of sight 
resulted to be $i \sim 30^\circ$ (Reis, Fabian, \& Young 2009).

In order to unambiguously resolve the line profile, we obtained a 
dedicated {\xmm} observation of this source that was performed in August
2006. However, {\xmm} caught the source during a hard state, and
the line was observed to be much weaker than during the Chandra
observation (these results will be presented in a forthcoming
paper). We therefore asked for a Target of Opportunity observation
with {\xmm} in order to observe the source during a soft
state. The ToO trigger was a source count rate higher than 10 c/s
observed by the All Sky Monitor (ASM) on board RXTE. This last observation 
was performed on August 24th 2008 (when the ASM count rate was about 20 c/s) 
for a total exposure time of about 45~ks. In this paper we present the 
spectral analysis of the data from this last {\xmm} observation.

The EPIC--pn camera was operated in timing mode, to prevent photon pile-up 
and to allow an analysis of the coherent and aperiodic timing
behaviour of the source. The EPIC--MOS cameras were switched off during the 
observation in order to allocate as much telemetry as possible to the pn,
and the Reflection Grating Spectrometer (RGS) was operated in the standard 
spectroscopy mode. 
Since we are interested here in a study of the iron line profile at 
$6.4-7$~keV, we focus our study only on EPIC--pn data.
Data were extracted and reduced using SAS v.8.0.0. We produced a calibrated
EPIC--pn event list through the \textit{epproc} pipeline. 
We extracted the source spectrum selecting a 13 pixels wide stripe around 
the source position equivalent to $53.3"$ (which should encircle more 
than $90\%$ of the energy up to 9 keV\footnote{see {\xmm} Users handbook, 
issue 2.6, available at 
http://xmm.esac.esa.int/external/xmm\_user\_support.}), 
and considered only PATTERN$\leq$4 and FLAG=0 events as a standard
procedure to eliminate spurious events. The background 
spectrum was extracted from a region of the pn field of view as far
as possible from the source position.

For spectral analysis the EPIC--pn energy channels were grouped by a 
factor of four in order to avoid an oversampling of the energy resolution
bin of the instrument. The X-ray spectral package we use to model the 
observed emission is HEASARC {\it XSPEC} v.12.3.0. 

The average count rate during our observation of {\1705} was 
767 c/s in the whole pn energy range. This is below the maximum count 
rate to avoid deteriorated response due to photon pile-up for pn 
observations in timing mode, that is 800 c/s as reported in the 
{\xmm} user handbook. 
However, in order to check that pile-up does not significantly affect 
the pn spectrum, we have also extracted the pn spectrum excluding the 
contributions from the central brightest pixels in order
to minimise the pile-up effect. Results from the two spectral 
sets are perfectly consistent with each other, and we therefore 
decided to proceed using the spectrum with the highest signal-to-noise
ratio.



We started to fit the continuum in the pn 2.4--11 keV energy range
with the typical model used for neutron star LMXBs of the atoll class,
which revealed to be the best fit continuum for this source too (see 
e.g.\ Di Salvo et al.\ 2005; Piraino et al.\ 2007; Barret \& Olive 2002), 
that is a blackbody and a thermal Comptonised component modelled by 
{\it comptt} (Titarchuk 1994), modified at low energy by photoelectric 
absorption modelled by {\it phabs} in XSPEC. This continuum model gave, 
however, an unacceptable fit, corresponding to a $\chi^2 / dof = 3862 / 422$ 
for the presence of evident localised residuals in the whole pn range. The
most prominent is a clear iron line profile at energies from 5.5 to
7.5 keV at more than $10\; \sigma$ from the best fit continuum 
(see Fig.~1).

\begin{table}
\caption{The best fit parameters of the continuum emission of the {\xmm}/pn spectrum 
of {\1705}. The blackbody luminosity is given in units of $L_{37} / D_{10}^2$,
where $L_{37}$ is the bolometric luminosity in units of $10^{37}$~ergs/s and 
$D_{10}$ the distance to the source in units of 10~kpc. The blackbody radius
is calculated in the hypothesis of spherical emission and for a distance of
7.4~kpc. 
Flux is calculated in the $2-10$ keV band. Uncertainties are given at 
$90\%$ confidence level.
}
\begin{tabular}{|lc|}
\hline
Parameter & Value   \\
\hline
$N_H$ ($\times 10^{22}$ cm$^{-2}$) & $1.8 \pm 0.2$ \\
$kT_{BB}$ (keV)                    & $0.55 \pm 0.01$ \\
L$_{BB}$ ($L_{37} / D_{10}^2$) 	   & $3.1 \pm 0.2$ \\
R$_{BB}$ (km)			   & $38 \pm 2$ \\
$kT_{seed}$ (keV)		   & $1.26 \pm 0.03$ \\
$kT_e$ (keV)			   & $4.8^{+1.7}_{-0.5}$ \\
$\tau$				   & $6.3 \pm 0.4$ \\
\hline
Flux ($10^{-9}$ ergs cm$^{-2}$ s$^{-1}$)  & $5.57 \pm 0.06$ \\
total $\chi^2$ (dof)		   & $473~(407)$ \\
\hline
\end{tabular}
\end{table}

\begin{table}
\caption{Best fit parameters of the discrete features of the {\xmm}/pn 
spectrum of {\1705}.
Rest frame energies for absorption edges are taken from the HULLAC code
database, while rest frame energies for emission lines are taken from 
Drake (1988). The emission line of Fe~XXV is a triplet; a 
rest-frame energy of 6.70~keV would correspond to the resonant
transition, while a rest-frame energy of 6.67--6.68~keV would 
correspond to the intercombination lines. The same holds for the
rest frame energies of Ca~XIX of the resonance (3.902 keV) and
intercombination (3.88--3.89 keV) transitions.
Uncertainties are given at $90\%$ confidence level.
}
\begin{tabular}{|lcc|}
\hline
Parameter & Value   & Identification \\
          &         & (Rest Frame Energy) \\
\hline

$E_{line}$ (keV) & $2.62^{+0.02}_{-0.01}$ & S~XVI  \\
$I_{line}$ ($10^{-3}$ ph cm$^{-2}$ s$^{-1}$) & $1.6 \pm 0.3$ & (2.6227 keV) \\
EqW (eV)         & $7 \pm 3$  & \\
		 &	      & \\
$E_{line}$ (keV) & $3.31 \pm 0.01$ & Ar~XVIII  \\
$I_{line}$ ($10^{-3}$ ph cm$^{-2}$ s$^{-1}$) & $2.0 \pm 0.2$ & (3.323 keV) \\
EqW (eV)         & $9 \pm 2$  & \\
		 &	      & \\
$E_{line}$ (keV) & $  3.90 \pm 0.01$ & Ca~XIX  \\
$I_{line}$ ($10^{-3}$ ph cm$^{-2}$ s$^{-1}$) & $1.9 \pm 0.2$ & (3.88--3.90 keV) \\
EqW (eV)         & $11 \pm 1$  & \\
		 &	      & \\
$E_{line}$ (keV) & $6.66 \pm 0.01$ & Fe XXV  \\
$I_{line}$ ($10^{-3}$ ph cm$^{-2}$ s$^{-1}$) & $3.6 \pm 0.1$ & (6.67--6.7 keV) \\
EqW (eV)         & $56 \pm 2$  & \\
		 &	      & \\
$Betor$          & $-2.27 \pm 0.08$ &     \\  
$R_{in}$ ($G M / c^2$)  & $ 14 \pm 2$ &  \\  
$R_{out}$ ($G M / c^2$) & $3300^{+1500}_{-900}$ & \\  
Incl (deg)       & $39 \pm 1$ &  \\  
		 &	      & \\
$E_{smedge}$ (keV) & $8.3 \pm 0.1$ & Fe~XXV \\
$\tau_{smedge}$ ($\times 10^{-2}$)   & $6^{+6}_{-2}$ & (8.828 keV) \\
Smearing Width (keV)	 & $0.7^{+0.8}_{-0.3}$ & \\

\hline
\end{tabular}
\end{table}

In order to fit these residuals we added to
our continuum model Gaussian emission lines at $\sim 2.6$ keV (S~XVI),
$\sim 3.3$ keV (Ar~XVIII), $\sim 3.9$ keV (Ca XIX), and $\sim 6.7$ keV
(Fe~XXV), respectively.
This gave a great improvement of the fit, corresponding to a 
$\chi^2 / dof = 639 / 410$. All these lines appear to be 
significantly broader than the pn energy resolution, with their Gaussian
$\sigma$ going from 120 eV for S XVI to 160 eV for Ar XVIII, to 170 eV for
Ca XIX, and to 260 eV for Fe XXV, respectively.
Note that a relatively broad ($\sigma \simeq 40$ eV) emission feature 
at $\sim 2.6$~keV and compatible with the Ly$\alpha$ transition of S~XVI 
was already significantly detected in the Chandra/HETG spectrum of this 
source (Di Salvo et al.\ 2005).  
The fit is further improved if we add an absorption edge at 8.47 keV,
most probably caused by photoelectric absorption of Fe XXV (this fit
gives $\chi^2 / dof = 528 / 408$).

Since the iron line profile visible in the residuals shown in 
Figure~1 clearly suggests a double peaked shape, typical of a 
relativistically smeared line produced by reflection of the primary
Comptonisation spectrum from an accretion disc, we substitute the 
Gaussian at 6.7 keV with a diskline profile, obtaining again a 
significant improvement of the fit ($\chi^2 / dof = 486 / 405$,
corresponding to a $\Delta \chi^2 = 42$ for the addition of three 
parameters). 
All the diskline parameters appear very well determined by the fit, 
and their best-fit values do not depend from the particular
choice of the continuum model.
The widths of the other Gaussian lines are also quite large,
$\sigma \ge 120$ eV, suggesting that these lines may also be affected 
by Doppler and relativistic distortions.  
To prove this hypothesis, we hence substitute these Gaussian lines with 
disklines, with all the parameters fixed at those of the iron
diskline, and we just fitted the line centroid energy and the normalisation 
of these components. In this way we obtain an equally good fit, with a 
$\chi^2 / dof = 491 / 408$ (note that the $\chi^2$ is higher but we have
now 3 free parameters less with respect to the fit with Gaussian lines).
We note that, although the uncertainties obtained from the fit are very
small (less than 1\% on the line energies), the rest-frame energies we 
obtain for all these low-energy lines are perfectly compatible with 
the rest-frame energies of the strongest expected transitions in the 
pn (2.4--10 keV) range (see Table~2). 
Moreover, from the line rest-frame energies we can infer that lighter 
elements (S and Ar) are in a H-like configuration, while heavier 
elements (Ca and Fe) are in a He-like configuration. This suggests a
ionisation parameter of $\log \xi \simeq 2.7-3$,
where $\xi$ is defined as $L_X / (n r^2)$, where $L_X$ is the ionising
X-ray luminosity, $n$ the density in the reflector, and $r^2$
the distance with respect to the emitting central sources.

\begin{figure}
\begin{center}
    \begin{minipage}{8cm}
      \includegraphics[width=8cm]{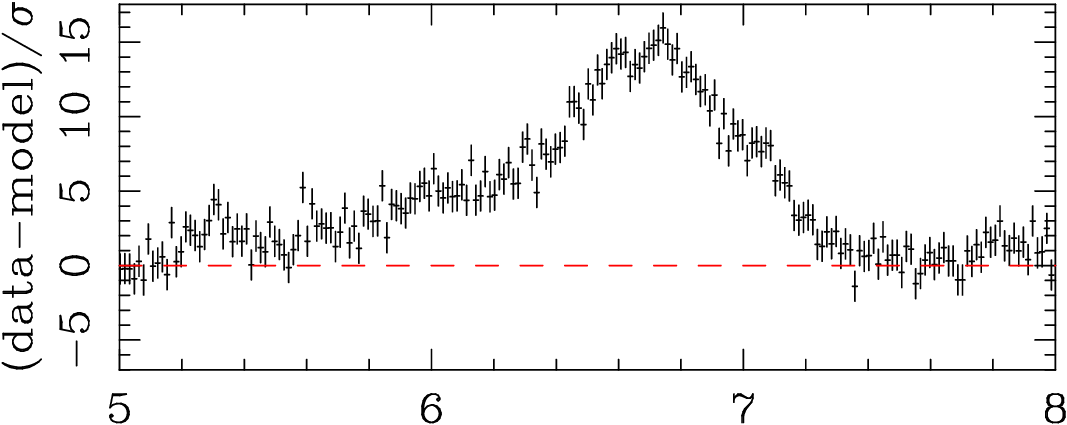}
\vskip 0.5cm
      \includegraphics[width=8cm]{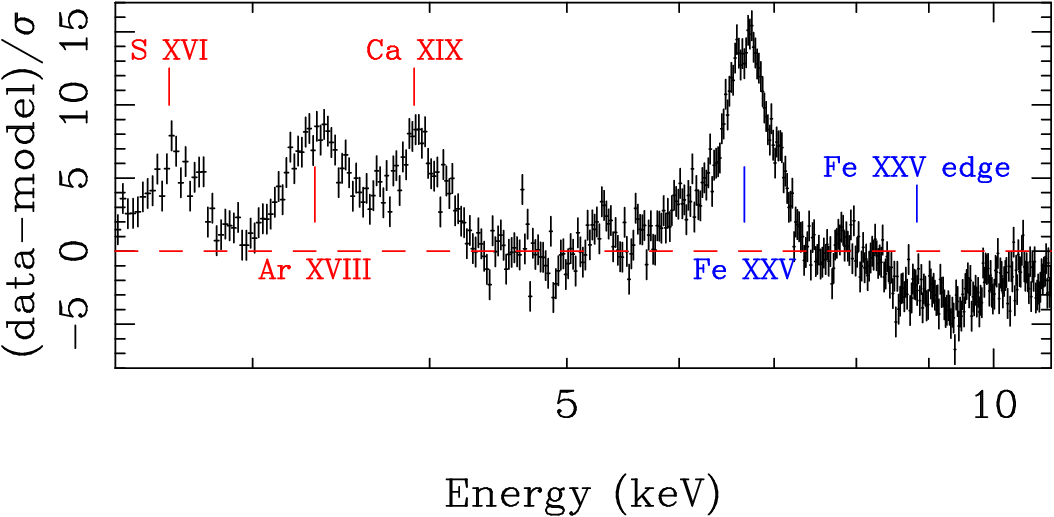}
    \caption{{\bf Top:} Residuals in units of $\sigma$ in the $5-8$ keV pn 
range with respect to the continuum model, composed by the photoelectric
absorption {\tt phabs}, a blackbody and the Comptonisation model 
{\tt comptt}, fitted in the whole pn range but excluding the  
$5.5-7.5$ keV range. The iron line profile, which appears to be broad and 
double peaked, is clearly visible in the residuals, at more than 
$10 \sigma$ above the continuum model.
{\bf Bottom:} Residuals in units of  $\sigma$ in the whole pn range with
respect to the best fit continuum model. Vertical lines are placed at 
2.62 keV, 3.32 keV, 3.90 keV, 6.67 keV, and 8.828 keV, respectively.}
    \label{fig1}
    \end{minipage}
\end{center}
\end{figure}
 
The fit is further improved if we substitute the absorption 
edge at 8.47 keV with a smeared edge, obtaining a final 
$\chi^2 / dof = 473 / 407$. This corresponds to a $\Delta \chi^2 = 18$
for the addition of 1 parameter; an F-test would give a probability of
$\sim 10^{-4}$ that the addition of the smearing improves the fit by
chance.
The best fit parameters are reported in Table 1 (continuum) and Table 2 
(discrete features), respectively; data and residuals, 
in units of $\sigma$, are shown in Fig.~2. In Fig.~3 we show the unfolded 
spectrum of the source.

Finally, to check the statistical significance of the diskline profile
for the iron emission line, we have substituted the diskline with a Gaussian
in order to compare the corresponding $\chi^2$s. In this case, the 
$\chi^2 / dof$ increases from $473 / 407$ using a diskline profile to
$537 / 410$ using a Gaussian profile for the iron line. The diskline
gives a better fit, with a $\Delta \chi^2 = 64$ for the addition of
3 parameters (an F-test would give a probability of chance improvement
of the fit of $3 \times 10^{-11}$). We also exclude that the observed line
profile may be a combination of several lines from iron in different 
ionisation states (as suggested by Pandel et al.\ 2008 for the case of 
4U~1636--536). In fact, using two Gaussian profiles to describe the iron
emission line, we find a $\chi^2 / dof = 494 / 407$ which is worse than
what we obtain using just one diskline profile. Moreover, the Gaussian
centroid energies are not easily explained in terms of blending of 
different iron lines, since the strongest Gaussian 
peaks at $6.70 \pm 0.01$~keV, while the other Gaussian (much weaker than 
the first one) is centred at $6.01 \pm 0.04$~keV and tries to fit the
red peak of the line profile. Fixing this line energy to $6.4$~keV (or
using two diskline profiles at 6.4 keV and 6.7 keV, respectively, 
instead of Gaussians) does not improve the fit.

\begin{figure}
\begin{center}
    \begin{minipage}{8.5cm}
      \includegraphics[width=5.7cm,angle=270]{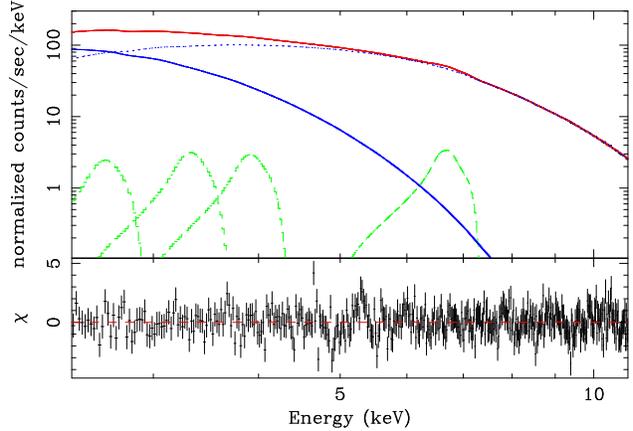}
    \caption{pn data (top) and residuals in units of $\sigma$ with respect to
the best fit model of Table~1 (bottom) of {\1705}. The single model components
({\tt comptt} indicated by the dotted line, {\tt blackbody} indicated by the thick 
solid line, and dashed {\tt disklines} for S~XVI, Ar XVIII, Ca XIX, and Fe~XXV, 
respectively) are also displayed. 
Note that the error bars on the single pn data points are so small that these 
points are covered by the solid line for the total model. }
    \label{fig2}
    \end{minipage}
\end{center}
\end{figure}
\begin{figure}
 \begin{center}
    \begin{minipage}{8.5cm}
      \includegraphics[width=6cm,angle=270]{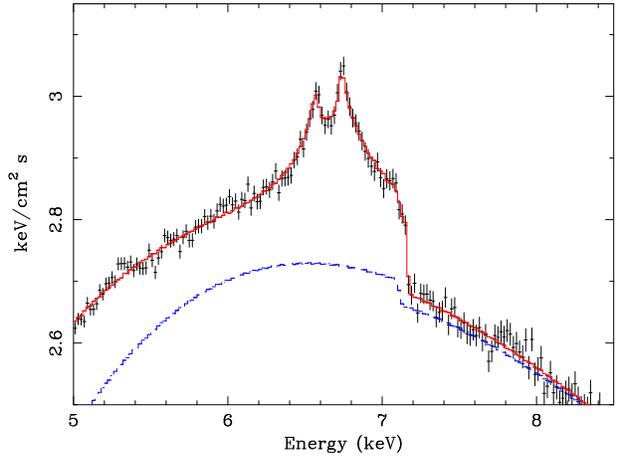}
    \caption{Unfolded pn spectrum of {\1705} in the $5-8.5$ keV range. The total 
model is indicated by the solid line on top the data points and the {\tt comptt} 
component is represented by the dashed line. }
    \label{fig3}
    \end{minipage}
\end{center}
\end{figure}

\section{Discussion}
The iron K$\alpha$ line at $6.4-7$ keV is one of the most important
diagnostic tools in X-ray spectra of accreting black holes.
In particular, if the line is produced in the innermost part of the
accretion disc, the line profile is distorted by Doppler effects
caused by the fast Keplerian motion and relativistic effects caused by
the distorted space-time very close to the compact object. These
effects produce characteristic smeared and asymmetric profiles, from
which we can derive essential information on the physical
parameters of the accretion disc close to the compact object. 
The same physics should apply to X-ray binaries containing a
low-magnetised neutron star, since the geometry of the accretion flow
close to the central compact object is very similar. Indeed, compelling
evidences have been found that iron line profiles in neutron star LMXBs
are smeared by relativistic effects in the inner accretion disc. The 
first clear example was the {\xmm} spectrum of atoll source Ser X--1 
(Bhattacharyya \& Strohmayer 2007), where the fit of the line profile with 
a Laor model was statistically preferred with respect to a simple Gaussian 
model. The relativistic nature of the iron line profile in Ser X--1 was
also confirmed by a Suzaku observation (Cackett et al.\ 2008). Cackett et 
al.\ (2008) also detect relativistic iron line profiles in Suzaku observations 
of the Z-source GX 349+2 (see also Iaria et al.\ 2009) and the atoll source 
4U~1820--30. A relativistic line profile is also observed in an {\xmm} 
observation of the Z-source GX~340+0 (D'A\'\i\ et al.\ 2009) and in the
atoll source and accreting millisecond pulsar SAX~J1808.4--3658 (Papitto 
et al.\ 2009; Cackett et al.\ 2009). A broad iron line profile is also
observed in the {\xmm} observation of the bright atoll 4U~1636--536
(Pandel et al.\ 2008); in this case the authors suggest that 
the broad line should be interpreted as a blending of several lines from 
iron in different ionisation states. Finally, the presence of a relativistic 
iron line profile in Chandra data of the atoll source {\1705} was suggested 
by Di Salvo et al.\ (2005) and recently confirmed by Suzaku observations 
of this source (Reis et al. 2009).


In this paper we have presented a high quality spectrum of the 
LMXB and atoll source {\1705} acquired by {\xmm} during a soft
state. We have shown that this spectrum is well described by a 
coherent disc-reflection scenario, confirming with the highest
statistics so far that the broadening mechanism is due to 
Doppler and relativistic effects at the inner edge of an 
accretion disc. The high statistics of the {\xmm}/pn spectrum allows
us to determine the best fit parameters with unprecedented
precision. The parameters of the continuum are in good agreement with
previous results (e.g. Piraino et al.\ 2007).  The presence of
several emission lines and an absorption edge from highly ionised 
elements (the most abundant emitters in the pn range) testifies 
the presence of a highly ionised matter in the reflector.
We have estimated a not-extreme value of the ionisation parameter, 
$\log \xi \simeq 2.7-3$, compatible with the presence of H-like S and
Ar and He-like Ca and Fe.

The iron line profile clearly shows the signature of a disc origin.
The rest-frame energy, $\sim 6.66$ keV, is compatible with the K$\alpha$ 
transition of He-like iron (Fe~XXV, resonance or intercombination
line), testifying that the reflecting matter is highly ionised.  
The line appears to be produced from $R_{in} = (14 \pm 2)\; R_g$ 
(where $R_g = G M/c^2$ is the Gravitational radius and it is 1.5 km for a 
$1\; M_\odot$ neutron star) to $R_{out} \simeq 3300\; R_g$. The inner 
radius of the disc therefore results to be $R_{in} = (29 \pm 4)\; m_{1.4}$~km, 
where $m_{1.4}$ is the neutron star mass in units of $1.4\;M_\odot$.  
Note that, for reasonable values of the neutron star mass, this measure
clearly indicate that the disc is truncated significantly far from the
neutron star surface.
Interestingly the radius of the soft blackbody emission, that we identify
(in agreement with Piraino et al.\ 2007) with optically thick 
emission from the accretion disc, gives a radius of $38 \pm 2$~km,
calculated under the simple hypothesis of spherical emission and for
a distance to the source of 7.4 kpc (Haberl \& Titarchuk 1995). Despite
the uncertainties in the emission geometry and distance to the source, 
the coincidence between the inner radius of the disc as calculated from 
the diskline and from the blackbody emission, is at least compelling.
This is the first time to our knowledge that such a large consistency 
is found among all the spectral components in a neutron star LMXB.
The power law dependence of the disc emissivity, $r^{betor}$, 
is parametrised by the index $betor$
which is $-(2.27 \pm 0.08)$, indicating that the disc emission is 
dominated by irradiation by a central source (see Fabian et al.\
1989). Finally, the disc inclination with respect to the line of sight is
also well determined, $i = (39 \pm 1)$ degrees. 

Signatures of a disc reflection origin are present not only in the iron
line profile, but also in the entire spectrum, and in particular in the
discrete features present at soft X-rays. These are produced by the
most abundant elements in the pn band, again highly ionised (H or 
He-like elements). In particular we find emission features from
S XVI, Ar XVIII, and Ca XIX, respectively, and all of them appear
quite broad, with a FWHM which steadily increases with line energy,
as expected in the hypothesis of Doppler broadening. Fitting these
lines with disklines, where we fix the emissivity index, the inner and
outer disc radii, and the inclination angle to the corresponding values
of the iron diskline profile, we obtain an equally good fit, indicating
that these lines are compatible with being produced in the same region
where the iron line is produced.

As expected, the iron absorption edge from Fe~XXV is also detected
in the spectrum. However, it also appears to be smeared (with a width
of about 0.7 keV) and at an energy of $8.3 \pm 0.1$ keV, which appears 
to be significantly redshifted with respect to the rest-frame energy of 
8.828~keV (according to the HULLAC code value) by $\sim 6\%$. A smeared 
and redshifted edge is also in agreement with the hypothesis that this
feature is produced in the inner accretion disc. For instance, in 
the hypothesis that this redshift is explained by gravitational 
redshift, the relation between the observed energy and the 
rest-frame energy may be written as 
$E_{obs} = E_{rf} (1+z)^{-1}$, where $(1+z)^{-1} = [1-2GM/(rc^2)]^{1/2}$,
from which we can derive the (averaged) radius of the emitting region,
$r = (17 \pm 2)\;R_g$, that is very close in radius 
to the region where the iron {\it diskline} is produced. 

In conclusion, the extraordinary high quality {\xmm} spectrum of {\1705} 
can be considered as a testbed for testing and developing high resolution 
reflection models. Attempts to fit this spectrum with available reflection 
models (see e.g.\ Ross \& Fabian 2005 which consider reflection models from
highly ionised media) are in progress; the results are already good and 
will be published in a forthcoming paper. However, none of the reflection 
models available to date include all the features we observe in this 
spectrum; in particular emission lines from Ar and Ca are still missing. 
Developing self-consistent and complete reflection models appears to be 
a very important task in view of the future, high resolution, 
X-ray missions, since with these spectra we are really looking at the 
innermost accretion disc.

\thanks{We thank all the {\xmm} team, and in particular Matteo Guainazzi, for 
supporting this ToO observation. }


\end{document}